\documentclass[journal=ancac3,manuscript=letter]{achemso}

\pdfoutput=1

\usepackage[version=3]{mhchem} 

\usepackage{xcolor}
\usepackage[normalem]{ulem}
\usepackage{dcolumn}
\usepackage{bm}

\usepackage{epstopdf}
\usepackage{graphicx}
\usepackage{amsmath,amssymb}
\usepackage{color,soul}

\author{Samuel Berweger}
\email{samuel.berweger@nist.gov}
\affiliation[NIST Boulder]
{Applied Physics Division, National Institute of Standards and Technology, Boulder, CO}

\author{Hanyu Zhang}
\affiliation[NREL]
{Materials and Chemical Science and Technology Directorate, National Renewable Energy Laboratory, Golden, CO}

\author{Prasana K. Sahoo}
\affiliation[USF]
{Department of Physics, University of South Florida, Tampa, FL}
\alsoaffiliation[IIT]
{Materials Science Centre, Indian Institute of Technology Kharagpur, India}

\author{Benjamin M. Kupp}
\affiliation[Penn State]
{The Pennsylvania State University, State College, PA}
\alsoaffiliation[NIST Boulder]
{Applied Physics Division, National Institute of Standards and Technology, Boulder, CO}

\author{Jeffrey L. Blackburn}
\affiliation[NREL]
{Materials and Chemical Science and Technology Directorate, National Renewable Energy Laboratory, Golden, CO}

\author{Elisa M. Miller}
\affiliation[NREL]
{Materials and Chemical Science and Technology Directorate, National Renewable Energy Laboratory, Golden, CO}

\author{T. Mitch Wallis}
\affiliation[NIST Boulder]
{Applied Physics Division, National Institute of Standards and Technology, Boulder, CO}

\author{Dmitri V. Voronine}
\affiliation[USF]
{Department of Physics, University of South Florida, Tampa, FL}

\author{Pavel Kabos}
\affiliation[NIST Boulder]
{Applied Physics Division, National Institute of Standards and Technology, Boulder, CO}

\author{Sanjini U. Nanayakkara}
\email{sanjini.nanayakkara@nrel.gov}
\affiliation[NREL]
{Materials and Chemical Science and Technology Directorate, National Renewable Energy Laboratory, Golden, CO}

\title{Spatially Resolved Persistent Photoconductivity in MoS$_2$--WS$_2$ Lateral Heterostructures}

\begin{document}

\begin{abstract}

The optical and electronic properties of 2D semiconductors are intrinsically linked via the strong interactions between optically excited bound species and free carriers.
Here we use near-field scanning microwave microscopy (SMM) to image spatial variations in photoconductivity in  MoS$_2$--WS$_2$ lateral multijunction heterostructures using photon energy-resolved narrowband illumination. 
We find that the onset of photoconductivity in individual domains corresponds to the optical absorption onset, confirming that the tightly bound excitons in transition metal dichalcogenides can nonetheless dissociate into free carriers.
These photogenerated carriers are most likely n-type and are seen to persist for up to days, and informed by finite element modeling we reveal that they can increase the carrier density by up to 200 times.
This persistent photoconductivity appears to be dominated by contributions from the multilayer MoS$_2$ domains, and we attribute the flake-wide response in part to charge transfer across the heterointerface.
Spatial correlation of our SMM imaging with photoluminescence (PL) mapping confirms the strong link between PL peak emission photon energy, PL intensity, and the local accumulated charge.
This work reveals the spatially and temporally complex optoelectronic response of these systems and cautions that properties measured during or after illumination may not reflect the true dark state of these materials but rather a metastable charged state.
\end{abstract}

While the properties of two-dimensional semiconductors typically mirror those of their bulk counterparts, their properties are augmented by effects arising from their reduced dimensionality \cite{splendiani10,mak10}, the lattice symmetry \cite{mak14}, and enhanced interactions with the environment.
For instance, the strong many-body interactions resulting from the quantum confinement give rise to excitons and other bound species  \cite{li18,mak12} whose binding energies significantly exceed the thermal energy $k_{\rm B} T$ at room temperature and are thus stable under ambient conditions.
The interaction energies of these bound excited species \cite{mitioglu12, li18}, their optical transition dipole moments, and the competition between radiative and nonradiative decay are strongly affected by the presence of free charge carriers \cite{lien19} and neutral bound species, as well as environmental influences such as surface adsorbates \cite{hu19,nan14} or chemical treatments \cite{yao18,tanoh19}.
The influence of free carriers on the optical properties are particularly strong since they underpin the competition between neutral excitons (X$^0$) and carrier-bound excitons (trions, X$^-$ and X$^+$) \cite{mak12,li18} and can drastically affect the photoluminescence quantum efficiency \cite{lien19}.
The optical and electronic properties of these materials are thus inextricably linked, and disentangling the two has remained challenging \cite{zhang17} in large part due to spatial inhomogeneities that occur on nanometer scales and large variations in sample properties due to processing and growth.

A comprehensive understanding of the optoelectronic properties therefore requires spatially resolved characterization of both the electronic and optical properties to address their interplay and the influence of spatial inhomogeneities.
Optical microscopy can readily obtain diffraction-limited spatial resolution of several 100s of nanometers and near-field implementations can achieve resolution as high as 10s of nanometers \cite{sahoo19,park16,bao15}.
Unfortunately, in addition to indirectly modifying electrical characteristics through local heating, optical absorption leads to excited states that can produce long-term photodoping in MoS$_2$ \cite{dibartolomeo17,wu15} that can persist for days \cite{cho14}.
Directly measuring the electronic properties of 2D materials by device transport is routine \cite{radisavljevic11,li15}, though this only yields device-average properties.
Spatially resolved electronic information is typically obtained using atomic force microscope (AFM)-based characterization such as Kelvin probe force microscopy (KPFM).
KPFM can quantify the local work function \cite{li15,sun19} and can be used to estimate carrier densities in functional devices \cite{dagan19}, but it does not directly measure the free carrier density and requires electrically contacted samples.

In this work, we perform a correlated study of the local electronic, optoelectronic, and optical properties of transition metal dichalcogenide (TMD) lateral heterostructures consisting of MoS$_2$--WS$_2$ multijunctions.
The nanoelectronic characterization is performed using near-field scanning microwave microscopy (SMM, also called scanning microwave impedance microscopy, sMIM).
SMM interacts directly with the local intrinsic as well as photoexcited \cite{tsai17,johnston18,chu18} free carriers in a sample and can characterize electrically isolated structures placed on insulating substrates with spatial resolution limited only by the tip apex radius.
This work is enabled by a spectrally narrow ($\approx$~10~nm bandwidth) continuously tunable optical excitation source that we combine with SMM to study the photon energy-dependent spatial distribution of photogenerated carriers as well as their long-term behavior. 
We find good correspondence between the photon energy at which photoconductivity emerges in the individual domains, and the expected absorption onset, which confirms that excitons can readily dissociate into free carriers despite their large binding energies.
The optoelectronic response is dominated by photoinduced n-type charging that is most pronounced in the multilayer MoS$_2$ regions and can lead to a localized increase in carrier density by up to a factor of 200.
This persistent photoconductivity also leads to uniform n-type conductivity throughout all domains of the flake which we attribute in part to charge transfer across the heterointerface.
Using spatially correlated photoluminescence (PL) microscopy mapping, we further show a clear correspondence between the persistent photoconductivity and the interplay of X$^0$- and X$^-$-emission that arises from the presence of free carriers.

\section*{Results}

\begin{figure}[tb]
	\includegraphics[width=.95\textwidth]{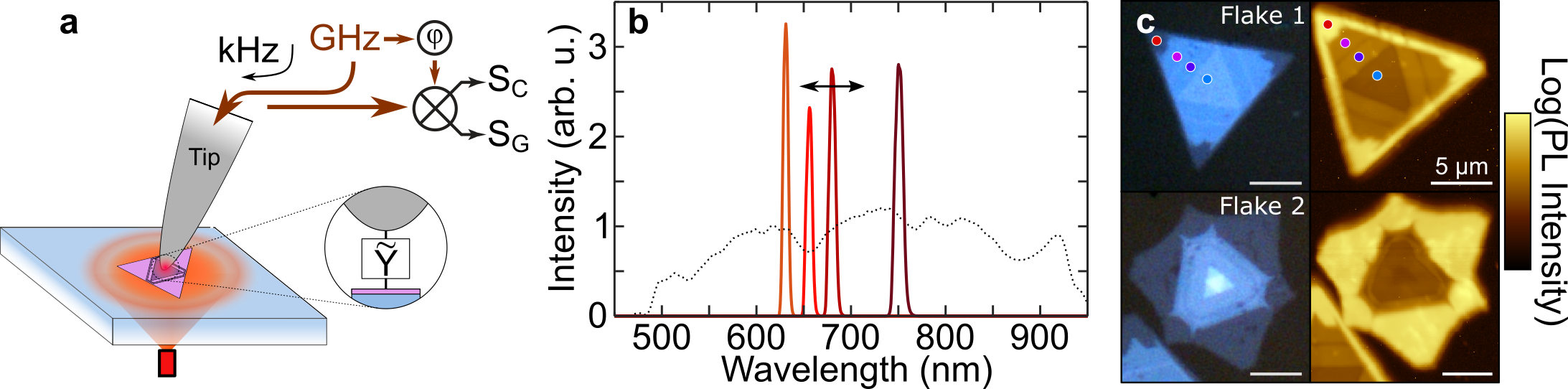}
	\caption{{\bf Optical microwave near-field microscopy of TMDs.}
		(a) Schematic of the experimental setup showing the uncollimated through-sample fiber illumination and tip-coupled microwave signal.
		(b) Sample narrowband excitation spectra generated from the broadband supercontinuum source (dashed line, not to scale).
		(c) Contrast-enhanced optical micrographs (left) and corresponding log-scale PL maps at an excitation photon energy of 2.38~eV (532~nm) showing the maximum PL emission intensity at each spatial pixel (right) of the two flakes studied in this work, denoted as flake 1 and flake 2.
		Different regions within the flakes can be identified based on emission intensity. 
	}
	\label{setup}
\end{figure}

Shown in Figure~\ref{setup}a is a schematic of the experimental setup. 
The SMM is based on a heavily modified AFM (Keysight / Agilent*) operating in contact mode \cite{berweger19}.
SMM is directly sensitive to changes in sample conductivity $\sigma$ via its effect on the complex-valued tip-sample admittance $\tilde{Y}=G+i\omega C$, with the conductance $G$, capacitance $C$, and frequency $\omega$.
Phase-resolved detection of the microwave signal $\tilde{S}$ is used to separate the real and imaginary terms, $S_{\rm G}$~$\propto$~$G$ and $S_{\rm C}$~$\propto$~$\omega C$, respectively.
Differential measurements \cite{berweger19} are performed by applying an AC tip bias with peak-to-peak amplitude $V_{\rm tip}$~=~1~V at a frequency $\nu_{\rm AC}$~=~50~kHz. 
Lock-in demodulation on $\tilde{S}$ at $\nu_{\rm AC}$ then yields the differential signals $S^\prime_{\rm G}$~$\equiv$~$dG/dV$ and $S^\prime_{\rm C}$~$\equiv$~$dC/dV$.

For SMM photoconductivity mapping, we use spectral amplitude shaping to generate a continuously tunable optical excitation source with $\approx$~10~nm spectral bandwidth from a broadband supercontinuum laser (Fianium Ltd*).
Shown in Figure~\ref{setup}b are representative narrowband excitation spectra at 630, 655, 680 and 750 nm (1.97, 1.89, 1.82, and 1.65 eV, respectively), as well as the full source supercontinuum spectrum (dashed line, not to scale).
Unless otherwise noted, all images are acquired under illumination at the photon energy indicated.

The samples were also characterized using a micro-photoluminescence (PL) setup operating under ambient conditions using an excitation wavelength of 532~nm (2.38~eV).
Shown in Figure~\ref{setup}c are contrast-enhanced optical images of the two flakes studied in this work (left, see supporting information for original images and contrast-optimized ones) as well as the corresponding spatially resolved log-scale photoluminescence (PL) maps (right) showing the maximum emission intensity at each spatial pixel.
These flakes will be referred to as flake 1 and flake 2 throughout the remaining text.
The control of single layer (SL) \cite{sahoo18} and bilayer (BL) \cite{sahoo19a} growth using our one-pot method is established, and a comparison with extensively characterized as-grown flakes (see supporting information) confirms that the outer regions with weak optical contrast but strong PL emission are SL, while the interior multilayer (ML) regions with stronger optical contrast and weak PL emission are BL with the exception of thicker regions at the center of flake 2.

\begin{figure}[tb]
	\includegraphics[width=.95\textwidth]{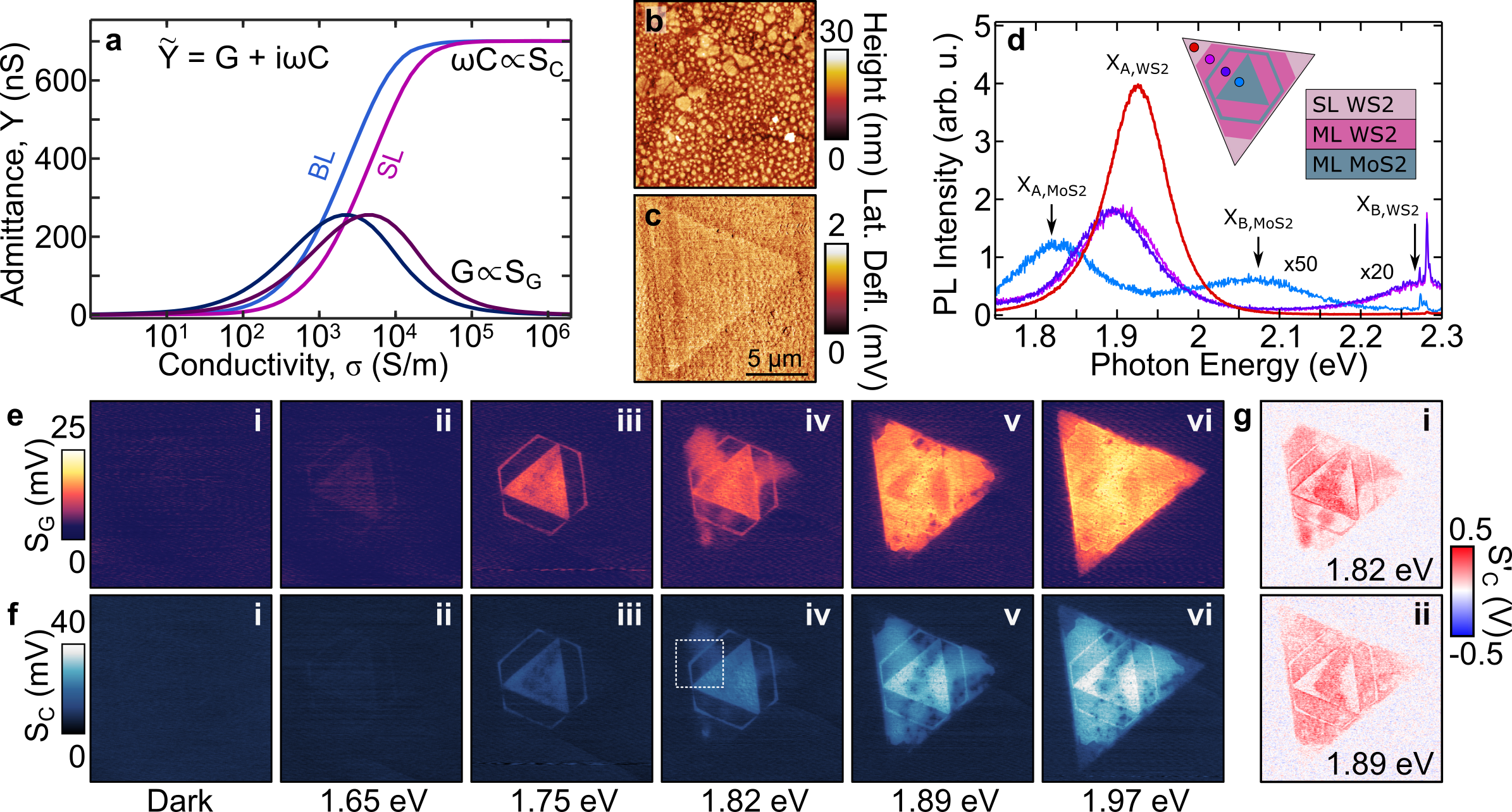}
	\caption{{\bf Photon energy-dependent heterostructure conductivity.}
		(a) Finite element modeling of the conductivity-dependent complex-valued tip-sample admittance for SL and BL TMDs as indicated.
		SL: Single layer, BL: Bilayer.
		(b) AFM topography of the position of flake 1, which can clearly be seen in (c) the corresponding AFM lateral deflection (friction) signal.
		(d) PL spectra from the corresponding  locations of flake 1 marked in Figure~1c and the inset, which schematically shows the different regions within flake 1.
		ML: Multilayer.
		The microwave near-field images of the S$_{\rm G}$ (e) and S$_{\rm C}$ (f) channels show the evolution of the spatial conductivity distribution with increasing excitation photon energy.
		(f) The $S^\prime_{\rm C}$ signal at photon energies of 1.82 and 1.89 eV show no change in sign between the MoS$_2$ and WS$_2$, indicating the same carrier type throughout. 
	}
	\label{sequence}
\end{figure}

SMM is directly sensitive to the local carrier density via the sample conductivity $\sigma_i$~=~$n_i e \mu_i$ with the elementary charge $e$ and the corresponding mobility and carrier density, $\mu_i$ and $n_i$, respectively ($i$~=~$h$, $e$).
Shown in Figure~\ref{sequence}a is the finite-element simulated complex-valued tip-sample admittance $\tilde{Y}$~=~$G+i\omega C$ at our chosen frequency of 17.2~GHz as a function of conductivity for SL and BL flakes as indicated.
For few-layer 2D materials, the admittance depends on the sheet resistance, a function of the layer thickness.
As the film thickness changes, the admittance curve is thus moved to higher (thinner) or lower (thicker) conductivity values.
Importantly, the capacitive signal (S$_C$) increases monotonically with sample conductivity, providing a qualitative measure of its spatial distribution.

Shown in Figure~\ref{sequence}b is the contact-mode AFM topography of flake 1.
Although the AFM used can readily resolve SL flakes \cite{berweger15}, the flake is obscured by substrate roughness.
The location of the flake is clearly seen in the lateral deflection signal shown in Figure~\ref{sequence}c, which arises from differences in friction as the tip is pulled across the sample.
The different domains of the heterostructure are schematically shown in the inset of Figure~\ref{sequence}d.
As confirmed by the lack of streaking artifacts in the AFM topography and lateral deflection, we ensure that the samples are free of polymer residue by repeated scanning.

Shown in Figure~\ref{sequence}d are PL spectra of the flakes acquired from the corresponding positions marked in Figure~\ref{setup}c and the inset, which schematically shows the different regions of flake 1.
MoS$_2$ and WS$_2$ have SL optical bandgaps of $\approx$~1.8 and $\approx$~2.0 eV \cite{splendiani10,zhu15} with the absorption blue-shifted relative to the PL by a few 10s of meV \cite{borys17}.
The strongest emission seen in Figure~\ref{setup}c originates from the direct bandgap A-exciton (X$_A$) emission of the SL domain of WS$_2$ \cite{zhu15} present on the outside of most flakes on this sample.
In contrast, the significantly diminished PL intensity in the interior of the flakes originates from ML domains of WS$_2$ and MoS$_2$ whose PL spectra show both X$_A$ and B-exciton (X$_B$) emission \cite{borys17,zhu15}.
Although the spectral resolution is optimized to measure the broad PL emission, the stokes-shifted Raman emission \cite{zhang15} is also seen at the position of the WS$_2$ X$_B$ emission in Figure~\ref{sequence}d.
We note that although the ML domains all show weak PL emission, the ML WS$_2$ shows consistently stronger emission compared to the ML MoS$_2$.

In Figure~\ref{sequence}e and f  we show the evolution of the S$_{\rm G}$ and topographically corrected \cite{coakley19} (see supporting information) S$_{\rm C}$ microwave near-field images in chronological order from left to right as the illumination is turned on and the photon energy increased.
Under dark conditions the conductivity in all domains of the flake is below the instrument sensitivity limit with no discernible contrast.
As discussed in detail below, charging and residual free carriers are seen for up to $\approx$~72 hours in these flakes after illumination; therefore the data shown in Figure~\ref{sequence}ei and fi was obtained after 4 days of illumination-free conditions to ensure that no photo-residual effects were measured.

While no discernible conductivity is seen in the uncharged dark state, after illumination of the flake with 1.65~eV excitation, weak photoconductivity emerges in the MoS$_2$ domains.
As the photon energy is increased to 1.75~eV and approaches the X$_A$ transition \cite{borys17}, the photoconductivity in the MoS$_2$ domain increases.
Upon excitation at the approximate X$_A$ transition energy at 1.82~eV, the photoconductivity in the MoS$_2$ domain further increases and a discernible conductivity is also seen over portions of the ML WS$_2$ domain.
At 1.89 eV the conductivity over the whole flake is seen to increase and a weak contrast emerges at the SL WS$_2$ vertices of the flake. 
Lastly, at a photon energy approximately corresponding to the WS$_2$ X$_A$ transition at 1.97~eV, all domains of the flake show discernible photoconductivity, including the SL WS$_2$ at the vertices and outer edges, though the higher contrast in the S$_{\rm C}$ channel suggests that the MoS$_2$ domains remain the most conductive.
Of particular note are spatially localized regions of reduced photoconductivity visible throughout the flake.
A careful comparison with the substrate topography suggests that these are not related to topographically-induced strain, though many appear to originate from dark regions in the optical images of Figure~\ref{setup}c (see supporting information for contrast-optimized images).

As differential measurements are directly sensitive to the local carrier type via the phase of the lock-in response \cite{berweger19}, they directly identify local variations in the carrier type and provide complementary information to the conductivity images.
Shown in Figure~\ref{sequence}g are S$^\prime_C$ images acquired at photon energies of 1.82 eV and 1.89 eV.
The sign of the lock-in signal remains positive throughout the flake, indicating that the conductivity is dominated by the same charge carrier type in both materials.
Understanding the observed photoconductive response of these flakes thus requires a detailed examination of the persistent photoconductivity as well as the effects of the interface.

\begin{figure}[!tb]
	\includegraphics[width=\textwidth]{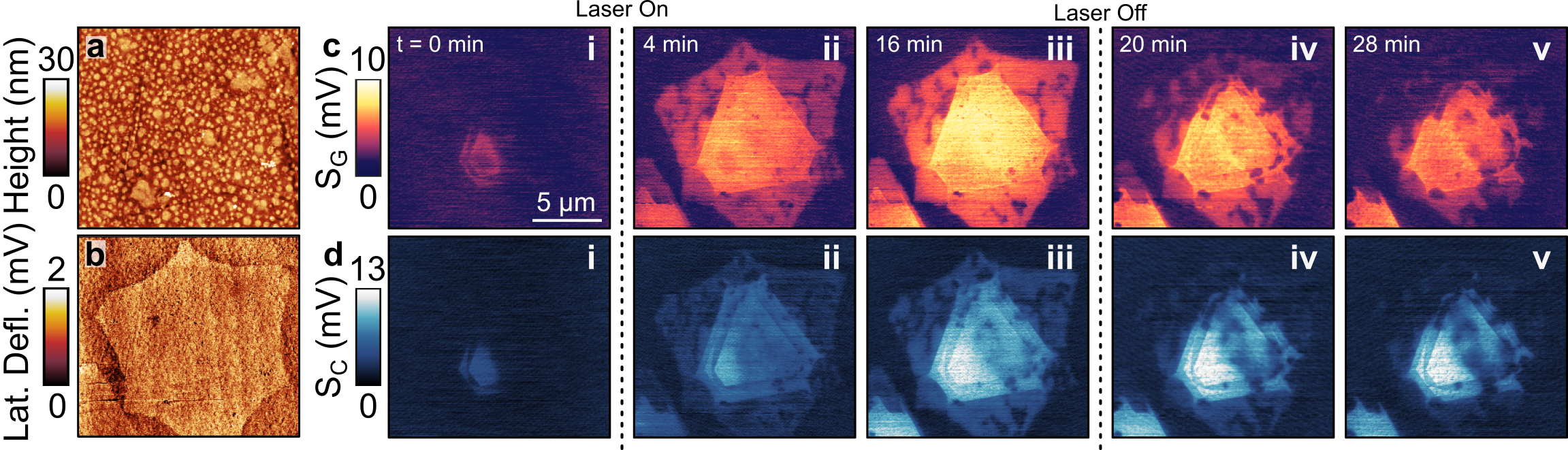}
	\caption{{\bf Photo-induced charging.}
		AFM topography (a) and AFM lateral deflection signal (b) of flake 2.
		A sequence of S$_{\rm G}$ (c) and S$_{\rm C}$ (d) images acquired from flake 2 at times from the first image as noted.
		Illumination was turned on and off during this sequence as indicated, showing the long-term kinetics associated with charging and discharging of these flakes.
	}
	\label{decay}
\end{figure}
We now turn to the dynamics of photogenerated charge carriers in these materials.
As noted above, long-term charging is observed in these flakes, with residual photoconductivity that is seen to persist for days.
Shown in Figure~\ref{decay}a and b are the AFM topography and lateral deflection signal of flake 2.
Shown in Figure~\ref{decay}c and d are select S$_G$ and S$_C$ images from the same flake (see supporting information for full data set).
Prior to illumination (t = 0), the flake is in a near-dark state.
The flake had been kept in dark conditions for 14 hours yet some residual photoconductivity is observed. 
In order for these flakes to revert to a nonconducting dark state such as that seen in Figure~\ref{sequence}ei, they typically have to be kept in dark conditions for $>$72 hours.
After the illumination with a photon energy of 1.97~eV is turned on, another scan is acquired that finished after 4 minutes.
All domains of the flake show a well-defined spatially inhomogeneous photoconductive response, with the strongest response seen in the ML domains and a weaker response in the SL WS$_2$ domain at the outside.

Under continued illumination, the photoconductivity also continues to increase as is evident from Figure~\ref{decay}ciii, acquired 16 minutes after the illumination is turned on.
As the flake charges the relative variations in photoconductivity are unchanged.
We then turn off the light immediately following the preceding image and see that a significant concentration of inhomogeneously distributed residual photoexcited carriers remains and decays slowly over time. 

Several noteworthy features stand out in the spatially inhomogeneous photoconductivity distribution. 
First, the interior ML MoS$_2$ domain that showed residual optically generated carriers even after several days in the dark shows the highest carrier density during charging as well as discharging, though this is highly spatially inhomogeneous across this domain.
This reflects the general observation across all flakes measured here (see supporting information for flake 1) that long-term (i.e., $\gtrsim$~12~hrs) charging occurs exclusively in the MoS$_2$ domain and the associated spatial inhomogeneity is highly reproducible.
Secondly, strong variations in photoconductivity are seen in the SL WS$_2$, with regions that do not show appreciable photoconductivity even after prolonged illumination.
Notably, the photoconductive decay appears to proceed outward from these regions after illumination is turned off.

These regions of reduced photoconductivity are a consistent feature of all flakes studied in this work (see supporting information for an additional flake).
The AFM topography and lateral deflection signal do not indicate any physical damage to the flakes themselves (i.e., tears or holes).
In some instances, these features appear to result from surface defects or adsorbates, which can result in strong local electronic effects as we previously observed \cite{berweger15}.
However, although many of these features appear to arise from dark regions in the optical images (Figure~\ref{setup}c) no apparent physical cause for these features is seen, suggesting that they may arise from local intrinsic defects introduced during growth or local transfer-induced strain.

We separately examined individual multilayer lateral heterojunction devices composed of only MoS$_2$ and WS$_2$ (see supporting information), where persistent photoconductivity was observed only in the MoS$_2$ species.
Importantly, no appreciable photoconductivity was observed for WS$_2$ even under resonant excitation, suggesting that photoinduced charging is necessary to obtain conductivity above the SMM sensitivity limit.
We also verified that hot-electron transfer from the tip \cite{tang18} or other scanning-related effects were not responsible for the observed sample charging (see supporting information).
The flake was illuminated with the tip retracted and the tip was engaged and used to scan only after the illumination was turned off, which nonetheless yielded an identical spatial distribution in residual charge and comparable discharging times.

\begin{figure}[tb]
	\includegraphics[width=.7\textwidth]{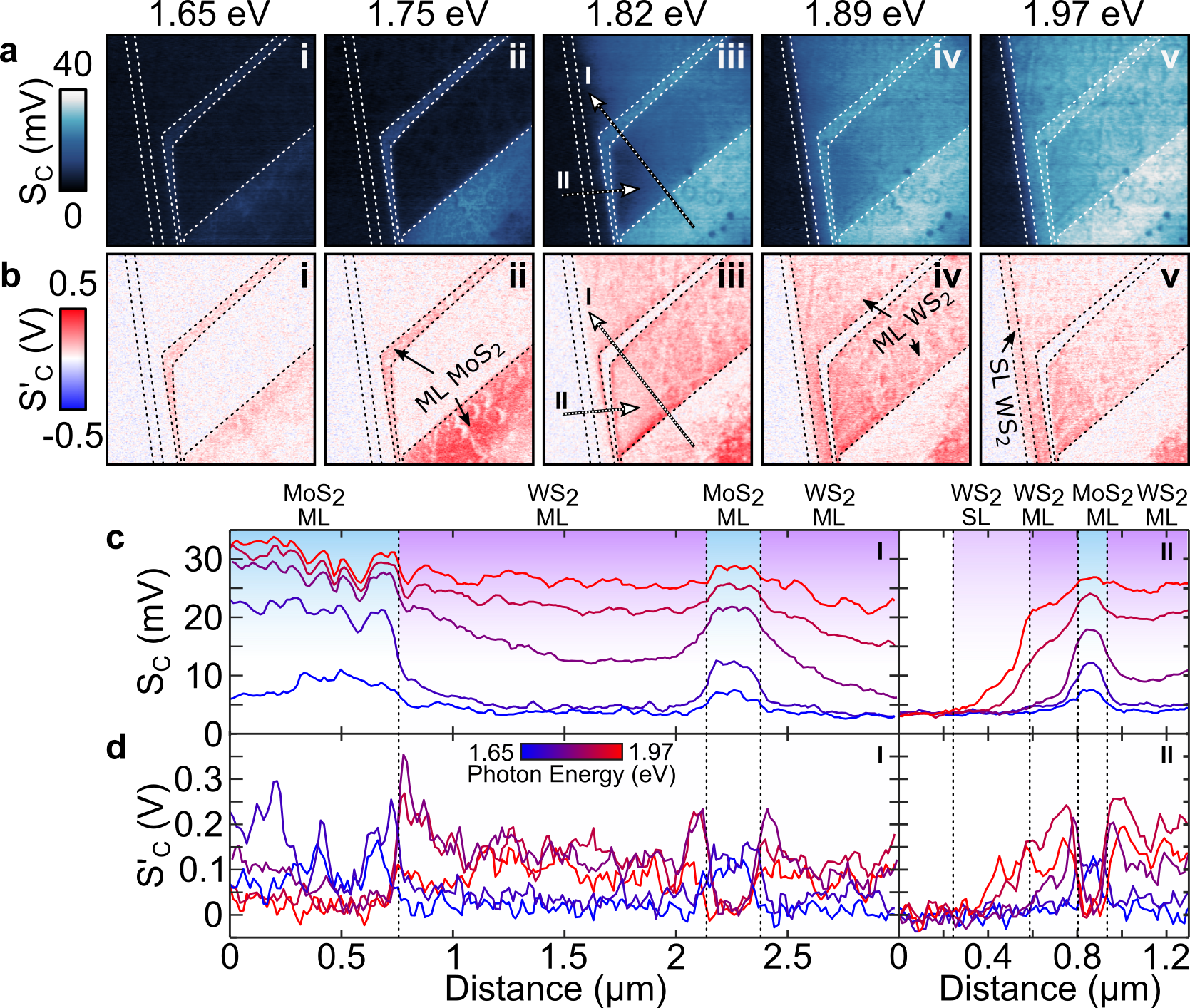}
	\caption{{\bf Examination of the heterointerface.}
		A photon energy-dependent sequence of S$_{\rm C}$ (a) and S$^\prime_{\rm C}$ (b) images taken from flake 1 at the location of the dashed square in Figure~\ref{sequence}fiv.
		Dashed lines indicate the positions of the boundaries between the individual domains, which are labeled in b.
		Line cuts along the dashed arrows shown in aiii and aiii are shown in (c) and (d).
	}
	\label{zoom}
\end{figure}

We now pivot to the question of the influence of the interface on the observed photoconductivity distribution.
Experimentally, photoinduced charge transfer across the MoS$_2$--WS$_2$ interface is often seen to generate a photovoltaic response \cite{gong14} -- including SL and BL devices fabricated from flakes grown by the same process as here \cite{sahoo18,sahoo19a} -- indicative of a type-II band alignment \cite{chen15,hill16}.
However, the band alignment may not be the same for lateral and vertical TMD heterostructures in general \cite{zhang18}, and for MoS$_2$--WS$_2$ in particular \cite{yang17}.
Furthermore, in part because the bandgaps differ by only $\approx$~200~meV with only a small band offset \cite{gong14}, the band alignment depends sensitively on strain and layer thickness \cite{kang15}, and may be further complicated by the presence of interfacial traps \cite{blackburn20} and alloying.
The nature of the interface can also be complicated by the possible presence of an interfacial exciton that can lead to increased recombination rates \cite{gong14}.

For examining the interface, we note that while we cannot directly probe the interfacial band alignment in these electrically isolated heterostructures, we can probe both the carrier density and type across the interfaces by examining high-resolution images of the interfacial regions.
Shown in Fig~\ref{zoom}a are the photon energy-dependent S$_C$ images of the region of flake 1 enclosed by the dashed square in Figure~\ref{sequence}fiv, with the domain boundaries indicated by dashed lines.
As expected from Figure~\ref{sequence}f, as the photon energy is increased the ML MoS$_2$ domain becomes photoconductive first, followed by the ML WS$_2$ domain, and lastly photoconductivity becomes discernible in the thin SL WS$_2$ domain at the flake edge at 1.97~eV.

The S$^\prime_C$ images shown in Figure~\ref{zoom}b provide complementary information to the S$_C$ images.
At low photon energies, we see that the S$^\prime_C$ signal increases in parallel with the S$_C$ signal.
As the photon energy is increased, regions of higher S$_C$ such as the ML MoS$_2$ domains show a diminished S$^\prime_C$ signal as expected for higher carrier densities \cite{huber12,edwards00}.
It should be noted that the  S$^\prime_C$ measurement probes the high-frequency branch of the capacitance-voltage (C-V) curve \cite{sze}. 
As such, any work function difference between our platinum tips and the doped sample does not invert the carrier type, but rather shifts the flat-band voltage while maintaining the sign of the C-V curve slope \cite{edwards00}.

Confirming the whole-flake S$^\prime_C$ image in Figure~\ref{sequence}f, the sign of the S$^\prime_C$ channel in Figure~\ref{zoom}b remains positive throughout, indicating that the conductivity in both materials remains dominated by the same carrier type.
Transport in MoS$_2$ is generally observed to be n-type in few-layer devices \cite{radisavljevic11,bao13,berweger15}, although this can be influenced by the presence of Schottky barriers at the contacts \cite{liu18}.
As further discussed below, photocharging in MoS$_2$ -- which occurs independent of electrode Schottky barriers -- is consistently observed to be n-type.
Previous transport measurements of the MoS$_2$ regions of identically grown flakes also showed n-type transport after control for Schottky barriers \cite{sahoo18,sahoo19a}.
Based on the S$^\prime_C$ signal we therefore conclude that the WS$_2$ is most likely n-type throughout these electrically isolated photoexcited flakes as well.

Careful examination of the interface suggests electron transfer from the MoS$_2$ to the WS$_2$.
Shown in Figure~\ref{zoom}c and d are the line cuts taken along the dashed arrows in Fig~\ref{zoom}biii where the approximate positions of the interfaces is estimated based on the photon energy-dependent photoconductivity onset.
In particular, the conductivity change across the ML MoS$_2$--WS$_2$ interface appears to be abrupt at lower (1.65~eV and 1.75~eV) and higher (1.97~eV) photon energies.
However, at 1.82~eV and to a lesser degree at 1.89~eV, the n-type conductivity in the ML WS$_2$ region is highest at the MoS$_2$--WS$_2$ interface and decays away from the interface.
This decay appears to occur over a length scale of 100~--~250~nm away from the interface, which is an order of magnitude shorter than measured depletion regions seen for MoS$_2$--WS$_2$ type-II alignment \cite{chen15} and indicates a different origin.
Given that this interfacial photoconductivity is observed to be strongest under resonant excitation of the ML MoS$_2$ domain at 1.82~eV where the WS$_2$ absorption is weak, this strongly suggests that these carriers originate in the MoS$_2$. 
This observed increased photoconductivity in the WS$_2$ region can be explained by the steady-state diffusion of electrons injected into the WS$_2$ from the MoS$_2$.

This charge transfer can occur by several possible mechanisms driven by the accumulation of electrons in the MoS$_2$. 
If the conduction bands are energetically close to each other (i.e., the band offset is shifted by a few 100~meV relative to a type-II junction), direct charge transfer could occur.
Alternatively, the decay of an exciton in the MoS$_2$ could readily impart sufficient energy on a free carrier to overcome interfacial barriers via an Auger-type mechanism.
In the case of type-II alignment, this charge transfer could also be indirectly mediated by interfacial excitons.
Here, an electron in the MoS$_2$ can bind with either an exciton or the hole of a dissociated exciton in the WS$_2$ to form an interfacial trion or exciton, respectively.
The decay of this interfacial species would then leave behind an excess electron in the WS$_2$ conduction band.
We note that the latter hypothesis is also supported by the observation that no interfacial conductivity is seen in the WS$_2$ at photon energies of 1.75~eV and lower despite the apparent accumulation of carriers in the MoS$_2$.
It thus seems likely that some absorption in the WS$_2$ (i.e., exciton generation) is necessary to drive this effect.
This mechanism could occur in any sufficiently sharp type-II junction supporting interfacial excited bound states such a excitons in van der Waals materials or charge transfer states in small molecule-TMD interfaces \cite{zhu18}.

\begin{figure}[!tb]
	\includegraphics[width=.95\textwidth]{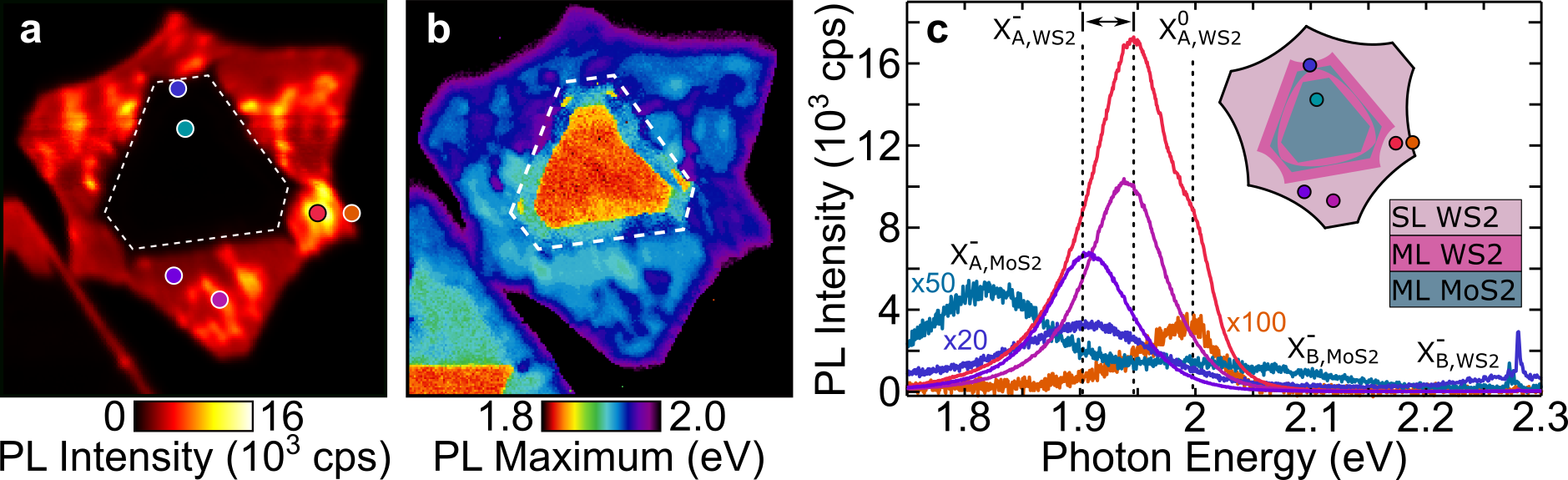}
	\caption{{\bf Photoluminescence mapping.}
		Spatially resolved PL intensity map of flake 2 at a photon energy of 1.97~eV (a) along with a map of the photon energy corresponding to the PL maximum at each spatial pixel (b).
		Dashed lines in a and b are a guide to the eye to discern the approximate position of the ML-SL WS$_2$ boundary.
		Select spectra are shown in (c) from the corresponding positions in a as well as the inset schematic of the flake, with the assigned spectral peak positions and the observed range of the WS$_2$ X$^-_{\rm A}$  emission indicated.
	}
	\label{emission}
\end{figure}

Having addressed the electronic characteristics and their spatial inhomogeneities in these samples under photoexcitation, we now directly compare these with spatially resolved PL maps.
Shown in Figure~\ref{emission}a is the spatially resolved PL map of flake 2 at an emission energy of 1.97~eV (630~nm), and the PL peak emission photon energy at each spatial location is shown in b.
A schematic of the different regions of the flake are shown in the inset of Figure~\ref{emission}c.
Together with the SMM images in Figure~\ref{decay} and the discrete PL spectra in Figure~\ref{emission}c that are taken from the indicated positions in Figure~\ref{emission}a and the inset, we find that our SL WS$_2$ domains contain three distinct regions.
The very edges of our flakes show asymmetrically broadened weak emission with a high-energy PL peak around 1.99~eV and very little discernible local conductivity.
This observed edge emission at 1.99~eV is the highest energy transition observed here, which we assign to the neutral A-exciton X$^0_{\rm A}$ in good agreement with literature values and corroborated by the low local photoconductivity \cite{mitioglu12,zhu15}.
In contrast, much of the WS$_2$ SL domain interior shows strong emission with a spectral maximum at around 1.91~eV and significant conductivity.
Lastly, we see localized regions within the SL domain that show emission up to 3x stronger than other interior regions and an emission maximum around 1.94~eV.
These strongly emitting regions correspond to the weakly photoconductive regions in Figure~\ref{decay}c and d.
Accordingly, the emission in the interior of the SL portion of the flake corresponds to the negatively charged trion X$^-_{\rm A}$.
The spatial variations in X$^-_{\rm A}$ show a correlation between a decrease in emission energy and increasing carrier density \cite{mak12}, with the low-conductivity regions revealing the superposition of X$^-_{\rm A}$ and X$^0_{\rm A}$ expected from regions nearing charge-neutrality \cite{zhu15}.

While the PL characteristics of the ML MoS$_2$ domain are uniformly weak and without discernible shift in the peak position, the ML WS$_2$ domains also contain regions with lower photoconductivity and a significant blueshift in the PL peak.
We also found no evidence of interfacial excitons to support our hypothesis of interfacial exciton-mediated charge transfer, though given the overall weak emission from the ML regions and the difference in excitation photon energy between the PL measurements (2.38~eV) and the interfacial conductivity observed with 1.82~eV excitation, this is not surprising.

\section*{Discussion}

Our observations directly confirm that within the SL WS$_2$ domains a locally reduced free carrier density leads to a blueshift in the PL peak and an increase in the PL efficiency \cite{nan14,lien19}.
Importantly, it is highly unlikely that the increased radiative decay rate is responsible for the reduced charging in the damaged regions.
Rather, the increased radiative decay rate is either the likely result of a low carrier density that results from passivated hole traps or the direct result of damage-related radiative recombination centers.
Any damage and related effects were likely incurred prior to SMM measurements, where we were careful to perform SMM characterization prior to PL measurements to minimize photoinduced damage.
We observed no photo-induced degradation \cite{kotsakidis19} during repeated measurement sequences over several weeks.

Persistent photoconductivity is a well-known semiconductor phenomenon and arises from defects that give rise to long-lived traps \cite{lang77}.
Persistent photoconductivity has been reported \cite{wu15,dibartolomeo17,cho14,lopez13} and deliberately engineered \cite{lee17} in MoS$_2$ devices.
This effect appears to consistently result in an accumulation of n-type carriers and is primarily driven by extrinsic interfacial traps at the MoS$_2$/SiO$_2$ interface \cite{wu15,dibartolomeo17,cho14}, and the disappearance of long-lived traps leading to hysteresis on hexagonal boron nitride (hBN) substrates \cite{lee13} further support this conclusion.
Although exfoliated and CVD-grown TMDs have intrinsic defect densities of up to 10$^{13}$~cm$^{-2}$ \cite{edelberg19,hong15} that adversely affect transport \cite{rhodes19}, these do not appear to directly produce long-lived traps, but rather facilitate their formation by providing chemically active sites for, e.g., oxygen adsorbates \cite{wu15}.
However, the role of these active sites on the otherwise chemically inert TMD surface in mediating the formation of extrinsic traps remains unclear.

We estimate the local carrier densities by noting that under illumination at 1.97~eV the interior ML domains in Figures.~\ref{sequence} and \ref{decay} show little signal variation in the S$_{\rm G}$ images although clear variations remain in the S$_{\rm C}$ images.
This suggests that the conductivity is at the plateau region of the S$_{\rm G}$ branch of Figure~\ref{sequence}A.
In contrast, under dark conditions the flake conductivity is uniformly at or below the instrumental detection limit, corresponding to a photoinduced increase in the conductivity by $\approx$~200 times.
Therefore, assuming carrier density-independent mobility and a carrier density near the detection limit, our finite element modeling indicates that at moderate optical intensities the carrier densities in regions of our structure is increased by a factor of up to $\approx$~200

As mobilities can be difficult to extract from two-terminal transport measurements and electrode deposition would likely damage our flakes, we assume a mobility of 5~--~10~cm$^2$/V$\cdot$s to estimate carrier densities, representative of CVD-grown MoS$_2$ \cite{rhodes19}.
For the interior BL regions, the photoexcited conductivity at the S$_{\rm G}$ plateau of $\approx$~2~$\times$~10$^3$~S/m corresponds to n~$\approx$~1.6~--~3.2~$\times$~10$^{12}$~cm$^{-2}$, while the detection limit of $<$~10~S/m corresponds to n~$<$~8~$\times$~10$^{9}$~cm$^{-2}$.
We note that this estimated dark carrier density is low given the high intrinsic defect density of these materials, suggesting that the dark conductivity is not significantly below the instrumental sensitivity limit.

This persistent charging appears to occur more broadly in MoS$_2$ and has significant implications for any measurement requiring prolonged optical illumination at photon energies at or above the onset of excitonic absorption - including below the electronic bandgap.
This implies that such measurements do not probe the intrinsic dark state but rather reflect a heavily modified metastable optically excited state.
Our results thus emphasize that optical measurements of MoS$_2$ and associated heterostructures that are sensitive to the free carrier density need be interpreted with caution and likely do not reflect the intrinsic sample itself.
However, the capability to develop spatially structured engineered trap storage devices \cite{lee17} aided by SMM imaging could also lead to novel memory and switching functionalities.

The large spatial variations in both PL peak position and intensity are characteristic of CVD-grown TMD sheets in general \cite{zhang17} and a range of observations are reported for SL WS$_2$ in particular \cite{kotsakidis19,hu19,yao18,gutierrez12}.
These results generally show a distinct PL intensity difference between the flake edge or near-edge region and the interior.
These variations are often directly correlated with a discernible shift in the emission peak that is attributed to spatial change in the X$^0_{\rm A}$ and X$^-_{\rm A}$ exciton emission due to local carrier density.
Here we can directly corroborate these assignments, which underscores the unique capability of SMM to directly access the local conductive and photoconductive behavior in electrically insulated samples in a nondestructive manner.

Lastly, we point out that despite the large exciton binding energies of up to several 100s of meV \cite{zhu15}, the free carriers that contribute to the microwave admittance are nonetheless generated \cite{tsai17}. 
This effect has been clearly established for low-dimensional systems supporting tightly-bound excitons using time-resolved microwave conductivity (TRMC) \cite{park15}.
And although the mechanism remains unclear, recent results from WS$_2$ flakes suggest a prominent role for ML regions in the formation of free carriers possibly due to reduced exciton binding energy and traps at lateral interfaces \cite{blackburn20}.

\section*{Conclusions}

In this work, we have used photon energy-resolved photoconductivity mapping to study the spatial photoconductivity distribution and long-term carrier accumulation in MoS$_2$--WS$_2$ lateral multijunction heterostructures.
We reveal that the optoelectronic response is dominated by photoinduced n-type charging that can increase the local carrier density by over two orders of magnitude and can persist for up to several days. 
This confirms that the tightly bound excitons in TMDs can nonetheless dissociate into free carriers, which challenges the assumption that optical excitation below the electronic bandgap does not significantly change the carrier density.
Through correlated photoluminescence mapping, we confirm that spatial variations in carrier density directly affect the emission intensity as well as the spectral peak position via the interplay between neutral excitons and carrier bound excitons (trions).
This work underscores the need for spatially resolved characterization in 2D systems and cautions that optical measurements -- particularly of 2D MoS$_2$ on SiO$_2$ -- can persistently modify the sample studied during the measurement process and likely reflect a metastable charged state rather than the true dark ground state.
Our results clearly demonstrate the potential of microwave near-field microscopy to probe the local variations in the potential energy landscape as it affects carrier motion, which has allowed us to address the interplay between optical and electronic properties.\\

\section*{Methods}

{\bf Heterostructure Sample.} 
MoS$_2$--WS$_2$ lateral multijunction heterostructures were grown using a one-pot chemical vapor deposition method described previously \cite{sahoo18,sahoo19a}.
On order to accommodate through-sample illumination, the samples were then transferred to an optically transparent quartz substrate using a solution-based poly(methyl methacrylate) (PMMA)--mediated transfer \cite{her13}.
Isolated flakes are chosen for study using an optical microscope.
MoS$_2$ and WS$_2$ domains within the flakes as well as single layer regions are confirmed using spatially resolved photoluminescence imaging using 532~nm (2.38~eV) illumination.

{\bf Optical Microwave Near-Field Microscopy.} 
The SMM is based on a commercial AFM/SMM (Agilent/Keysight 5500*).
As described previously \cite{berweger19}, the system is modified to accommodate external IQ mixer-based microwave demodulation.
All measurements are performed with the AFM operating in contact mode using SMM-specific solid Pt cantilevers (Rocky Mountain Nanotechnology 25Pt300A*).

The optical excitation is based on a broadband supercontinuum source (Fianium*). 
The continuously tunable excitation source is generated from the broadband supercontinuum spectrum by spectral shaping using a razor blade slit mounted on a translation stage and placed at the shaper Fourier plane.
After spectral shaping, the light is coupled into a multimode fiber and delivered to the sample by illumination through the substrate from below.
In order to provide uniform optical excitation across the tip-scanned region and maintain linear excitation densities, the fiber termination is placed immediately below the transparent substrate and the beam allowed to diverge.
This produces an excitation spot at the sample with radius $r$~$>$~100~$\mu$m and intensity $<$2~W/cm$^2$.

{\bf Finite-Element Modeling.}
The finite-element modeling was performed using COMSOL 4.4* as previously described in the literature\cite{kundhikanjana09,berweger15,tsai17}.
Briefly, the we use the AC/DC module with an axisymmetric geometry at our experimental frequency of 17.2~GHz.
The experimental geometry is modeled with a TMD flake ($\epsilon$~=~7) with thickness of 0.65~nm (SL) or 1.3~nm (BL) and variable conductivity on a quartz substrate ($\epsilon$~=~3.9).
We use a parabolic tip shape with an apex radius of 20~nm that is held 1~nm above the substrate to facilitate meshing.
\\
\\
*Mention of commercial products is for informational purposes only, it does not imply NIST's recommendation or endorsement.

\section*{Supporting Information Available}
Optical micrographs of lateral heterostructures (Figure S1); Optical and PL maps of flakes 1 and 2 (Figure S2); Photon-energy dependent conductivity of an additional flake (Figure S3); Correction of capacitive topographic cross-talk (Figure S4); Full photo-induced charging and discharging dataset for flake 2 (Figure S5); Decay of persistent photoconductivity of flake 1 (Figure S6); Verification of lack of charge transfer from tip (Figure S7); Photoconductivity of individual MoS$_2$ and WS$_2$ lateral heterojunctions (Figure S8).

\section*{Acknowledgements}

We would like to acknowledge valuable discussions with Nicholas J. Borys and Obadiah G. Reid.
HZ, JLB, EMM, and SUN are employees of the Alliance for Sustainable Energy, LLC, the manager and operator of the National Renewable Energy Laboratory for the U.S. Department of Energy under Contract No. DE-AC36-08GO28308. 
Funding provided by U.S Department of Energy, Office of Science, Office of Basic Energy Sciences, Division of Chemical Sciences, Geosciences and Biosciences.  
The views expressed in the article do not necessarily represent the views of the Department of Energy or the U.S. Government.
Contributions of the US Government are not subject to copyright in the United States.

\bibliography{lat_hetero} 

\end{document}